\documentstyle [prd,aps,psfig] {revtex}
\draft
\newcommand{\ang}{{\rm \AA}}
\newcommand{\be}{\begin {equation}}
\newcommand{\ee}{\end {equation}}
\newcommand{\bn}{\begin {eqnarray}}
\newcommand{\en}{\end {eqnarray}}
\newcommand{\bd}{\begin {displaymath}}
\newcommand{\ed}{\end {displaymath}}

\def\Ref#1{(\ref{#1})}

\begin {document}
\title {Time-scale invariance of relaxation processes of density fluctuation 
in slow neutron scattering in liquid cesium}
\author{R.M. Yulmetyev${}^{1,}$\thanks{e-mail:rmy@dtp.ksu.ras.ru},
A.V. Mokshin${}^{1,}$\thanks{e-mail:mav@dtp.ksu.ras.ru}, P. H\"anggi${}^{2}$ and
V.Yu. Shurygin${}^{3}$}
\address{$^{1}$Department of Physics, Kazan State
Pedagogical University, Kazan, Mezhlauk 1, 420021
Russia \\
$^{2}$Department of Physics, University of Augsburg,
Universit\"atsstrasse 1, D-86135 Augsburg, Germany \\
$^{3}$Department of Physics, Yelabuga State Pedagogical
Institute, Elabuga, Kazanskay 89, 423630 Russia}
\maketitle
\begin{abstract}
The realization of idea of time-scale invariance for relaxation processes
in liquids has been performed by the 
memory functions formalism. The best agreement with experimental data for the
dynamic structure factor $S(k,\omega)$ of liquid cesium near melting 
point in the range of wave vectors ($0.4 \ang^{-1} \leq k \leq 2.55
\ang^{-1}$) is found
with the assumption of concurrence of relaxation scales for memory functions
of third and fourth orders. Spatial dispersion  of the four first
points in spectrum of statistical parameter of non-Markovity $ \epsilon_{i}(k,
\omega)$ at $i=1,2,3,4$ has allowed to reveal the non-Markov nature of
collective excitations in liquid cesium, connected with long-range memory
effect.  
\end {abstract}  
The dynamic structure factor $S(k,\omega)$ of liquid metals (lithium,
sodium, rubidium, lead, cesium, aluminum, potassium) represents  unique
information on the collective excitations in these systems (see Ref.
\cite{Scop2}). Most of  existing theories for $S(k,\omega)$ are based  on
models of the linearized and generalized hydrodynamics \cite
{Egelstaff}. Despite much effort and considerable recent
progress,understanding of experimental data for liquid metals remains an
interesting challenge.
Here  
adequate understanding and explanation of high-frequency collective
excitations  does not exist. The reason is that the of nature such 
excitations in liquid
metals
is not described by any of hydrodynamic models. On account of specific
short-range and oscillatory behavior of ion-ion potential in liquid
metals, the hydrodynamics here is not applicable. 

In the present work, for an explanation of a specific nature of high-frequency
collective excitations in liquid cesium \cite{Boden.} we use one of most
fundamental ideas of modern physics - the idea of  invariance. In
particular, we suggest the idea of time-scale invariance of relaxation
processes in liquids. Experimental data on $S(k,\omega)$
give direct information about relaxation processes of density
fluctuation in
liquids. Under the assumption, for any experimentally observable
relaxation process there corresponds a multilevel hierarchy of 
interconnected relaxation processes. Actually, in experiment
the "top" of its relaxation "iceberg" is observed only. On a certain
relaxation level, an invariance (equiscaling) of two 
nearest  interconnected relaxation processes can exist . Such
 invariance can be easily taken into account by the memory functions 
formalism, which   most adequately describes nonequilibrium
statistical processes in condensed matter physics. 

Later we consider a normalized time correlation function (TCF)
$a(t)=< \delta A^{*}(0) \delta A(t)>/< \mid \delta A(0) \mid^{2}>$
of particle-density fluctuation in liquid metal $\delta A(t)=A(t)-<A(t)>$,
where $\delta A(t)=N^{-\frac{1}{2}} \sum_{j=1}^{N} exp(ikr_{j})$, $k$ is wave
vector and $r_{j}$ defines the position of the $j$th particle of the liquid. 
With the help of Zwanzig and Mori's projection operators method
\cite{Mori,Zwanzig} it is possible to construct the  infinite 
chain of connected non-Markov kinetic equations as follows
\bn
\label {chain}
\frac{dM_{i}(t)}{dt}&=&-\Omega_{i+1}^{2}\int_{0}^{t}d\tau
M_{i+1}(t-\tau)M_{i}(\tau),  i=0,1,... ,
\en
where $M_{0}(t)=a(t)$.
Here $M_{i}(t)$ is a memory function of $i$th order and $\Omega_{i}^{2}$
are general relaxation parameters with  the 
dimension of the square of frequency. These parameters are connected with
even frequency moments of spectral density of TCF $a(t)$ by the
following relationship \cite {Shurygin1} $\Omega_{1}^{2}=I_{2}$,
$\Omega_{2}^{2}=I_{4}I_{2}^{-1}-I_{2}$, 
$\Omega_{3}^{2}=(I_{6}I_{2}-I_{4}^{2})/(I_{4}I_{2}-I_{2}^{3})$.

Every relaxation process can be described with the help of a characteristic
time scale usually named as relaxation time.
So, for example, the relaxation time for initial TCF $a(t)$ can be determined
as follows $\tau_{a}= Re \int_{0}^{ \infty}dt a(t)$. Similarly,
relaxation time on the second relaxation level [at $i=1$ in 
Eq. \Ref{chain}] for memory function $M_{1}(t)$ would be 
expressible as
$\tau_{1}= Re \int_{0}^{ \infty}dt M_{1}(t)$,
where the symbol $Re$ means the real part.
 It is
convenient to describe long-range memory effects in the underlying system with
the help of time scales $ \tau_{a}$ and $ \tau_{M_{1}}$. For
example,  the presentation of the dimensionless non-Markovity parameter was
introduced earlier in Ref. \cite {Shurygin1}, 
$\epsilon_{1}=\tau_{0}/\tau_{1}$, as an criterion of describing of
non-Markovity for any relaxation 
processes .
As pointed out in Ref. \cite{Shurygin1} values of $\epsilon_{1}$
allow to obtain a quantitative and qualitative estimate of
non-Markovity effects and statistical memory in relaxation
processes. Parameter $\epsilon_{1}$ allows to divide  all relaxation
processes into three important cases. Markovian processes correspond to
$\epsilon_{1} \to \infty$, while quasi-Markovian processes are appropriate
in situations
with $\epsilon_{1} \gg 1$ and $\epsilon_{1}>1$. The limiting case
$\epsilon_{1} \sim 1$ 
describes non-Markovian processes. In this case the time scale of memory
processes and correlations (or junior and senior memory functions)
coincide with each other. Thus,  the introduced 
parameter $\epsilon_{1}$ characterizes non-Markovity and memory effects 
for any relaxation processes. 

The  infinite set of values of parameters 
$ \epsilon_{i}$, where $i=1,2,3,...$, was also entered on the
basis of the simple formula
$\epsilon_{i}= \tau_{i-1}/\tau_{i}$, 
where $\tau_{i}$ is a relaxation time of the memory function of
$i$th order \cite{Shurygin2}. 
The  whole set of values of non-Markovity parameter $\epsilon_{i}$ forms
the statistical spectrum, which is connected with collective and statistical
properties of the system and allows also to estimate in detail the non-Markovian 
properties of the underlying relaxation process.
In Ref. \cite{Yulmetyev1}  the conception of non-Markovity
parameter for the frequency-dependent case was generalized . This  parameter
is determined by the following expression
$\epsilon_{i}(\omega)=\left \lbrace \mu_{i-1}(\omega)/\mu_{i}(\omega) 
\right \rbrace^{\frac{1}{2}}$ at the fixed 
wave vector $k$ .
Here $i=1,2,3...$ and $ \mu_{i}(\omega)$ is the power frequency spectrum
of the $i$th relaxation level, which is defined in the following  way
$\mu_{j}(\omega)= \left \lbrack Re \int_{0}^{ \infty} dt e^{i \omega t} 
M_{j}(t) 
\right \rbrack^{2}$.
 
The appropriate values of $ \epsilon_{i}(\omega)$ and $ \epsilon_{i}$
will correspond to every $j$th equation of the chain \Ref{chain} . As
the chain of the Eqs. \Ref{chain} is infinite, the sets of
$\epsilon_{i}$ and 
$\epsilon_{i}(\omega)$ are also infinite.
Now let us use the approximation $M_{i+1}(t) \approx M_{i}(t)$.
It means the approximate equality of relaxation  time  scales
of memory functions of $i$th and $(i+1)$th  orders, i.e., $ \tau_{i+1}
\approx \tau_{i}$. Then the non-Markovity parameter for the whole  frequency range     
is approximately equal to unity $\epsilon_{i+1}(\omega) \sim 1$ (and also
$\epsilon_{i+1} \sim 1 $), and chain \Ref{chain} becomes closed.
Applying Laplace transform  to the $i$th equation of the chain \Ref{chain}  
we get
$\tilde M_{i}(s)= \lbrace -s+(s^2+4 \Omega_{i+1}^{2})^{\frac{1}{2}}
\rbrace / 2\Omega_{i+1}^{2}$. 

The results of experiment for liquid cesium  were published in Ref.
\cite {Boden.}. The dynamic structure factor  $S(k, \omega)$ has been
measured by inelastic slow neutron scattering (INS) near the melting point at $T=380$ K. The wave vector $k$
changes in range from $0.2$  
$\ang^{-1}$ up to $2.55$ $\ang^{-1}$. The numerical results
for $S(k)$ were obtained from experimental data \cite {Boden.} too.

It is well known that $S(k, \omega)$  is connected with TCF of density
fluctuation in the following way:
$S(k, \omega)= \lbrace S(k)/pi \rbrace \lim_{\epsilon \to +0} Re [\tilde a(k,i
\omega+ \epsilon)]$,
where Laplace transform $\tilde a(k,s)=\int_{0}^{ \infty}dte^{-st}a(k,t)$
was found by us as follows. We have 
taken advantage of the correlation approximation for the fourth order memory
function \cite{Our}
\be
M_{4}(t) \approx M_{3}(t).
\label {eq.21}
\ee 

From the physical point of view it means the concurrence of time scales of
TCF's $M_{3}(t)$ and $M_{4}(t)$. In this case the  chain of the connected
kinetic equations \Ref{chain} becomes a system consisting of four equations.
Using  Laplace  transform, we receive the following
equation for the dynamic structure factor $S(k, \omega)$:
\bn 
\label {eq.22}
S(k, \omega)&=& \frac{S(k)}{2 \pi} \Omega_{1}^{2} \Omega_{2}^{2} \Omega_{3}^{2}
(4 \Omega_{4}^{2}- \omega^{2})^{ \frac{1}{2}} \lbrace \Omega_{1}^{4}
\Omega_{3}^{4}+ 
                                                       \nonumber \\
& &\omega^{2}(-2 \Omega_{1}^{2} \Omega_{3}^{4}+
\Omega_{1}^{4} \Omega_{4}^{2}- \Omega_{1}^{4} \Omega_{3}^{2}
+2 \Omega_{1}^{2} \Omega_{2}^{2} \Omega_{4}^{2}-
\Omega_{1}^{2} \Omega_{2}^{2} \Omega_{3}^{2}+
\Omega_{2}^{4} \Omega_{4}^{2})+                                            
                                                     \nonumber \\
& &\omega^{4}( \Omega_{3}^{4}-2 \Omega_{1}^{2}
\Omega_{4}^{2}+2 \Omega_{1}^{2} \Omega_{3}^{2}-2
\Omega_{2}^{2} \Omega_{4}^{2}+ \Omega_{2}^{2} \Omega_{3}^{2})+ 
\omega^{6}( \Omega_{4}^{2}- \Omega_{3}^{2})
\rbrace^{-1}.
\en

Relaxation frequency parameters $\Omega_{1}^{2}$ and $\Omega_{2}^{2}$
are defined as follows \cite{Metallofizika}
\bn
\label {eq.23}
\Omega_{1}^{2}&=& K_{B} T k^{2} \cdot \lbrack m S(k) \rbrack^{-1}; \ \ \omega_{1}^{2}=
\Omega_{1}^{2};\ \ 
\Omega_{2}^{2}=\omega_{2}^{2}- \omega_{1}^{2}; \nonumber \\ 
\omega_{2}^{2}&=&3 \omega_{1}^{2}S(k)+ \omega_{l}^{2};\ \
\omega_{l}^{2}= N/\lbrack mV \rbrack \int drg(r)[1-\cos (kr)]
\bigtriangledown_{z}^{2} u(r).
\en

In Eqs. \Ref{eq.23} the following designations are introduced: $K_{B}T$
is thermal energy, $g(r)$ is a radial distribution function 
of particles, $u(r)$ is a pair interparticle potential of interaction,
 and axis $z$ is
chosen in the direction of a wave vector $k$. Calculating  the  frequency
parameter $\Omega_{2}^{2}$ we use the well-known approximation  \cite{Hubbard}
$\Omega_{2}^{2}=3 \Omega_{1}^{2}S(k)+ \omega_{E}^{2} \lbrace 1- \lbrack 3 \sin
(kR_{0})/kR_{0} \rbrack- \lbrack 6 \cos(kR_{0})/(kR_{0})^{2}\rbrack+
\lbrack 6 \sin 
(kR_{0})/(kR_{0})^{3}\rbrack \rbrace-\Omega_{1}^{2}$,
where $ \omega_{E}$ is the Einstein frequency. In our case
$ \omega_{E}$ has the following value $ \omega_{E} =4.12$ $ \cdot
10^{12}$ $s^{-1}$ \cite{Boden.}. This frequency parameters can also be calculated
through  $I_{i}$ $(i=2,4)$.

Theoretical  formulas for calculation of relaxation frequency
parameters $\Omega_{3}^{2}$ and $\Omega_{4}^{2}$ are also known
\cite{Metallofizika}. However the 
final result of these
calculations contains some errors. For example, in paper
\cite{Metallofizika}  it is shown that results of calculated
frequency moments in papers \cite{Michler,Dubey,Ebbsjo}, vary from $10$ \%  to 
$50$  \%, and the distinction reaches up to 30 times for separate
values of the wave vector. Therefore it is more convenient to obtain
these parameters by comparing   results of theory and experiment.
Relaxation
frequency parameter $ \Omega_{4}^{2}$ is easy to find from comparison of
developed theory with experiment on zero frequency. Namely, it follows
$\Omega_{4}^{2}= 
\pi^{2} \Omega_{1}^{4} \Omega_{3}^{4} 
\lbrace S(k,0) \rbrace^{2}/\lbrace \lbrack S(k)\rbrack^{2} \Omega_{2}^{4}
\rbrace^{-1}$ from
Eq.\Ref{eq.22} at $ \omega \to 0$. 

The spectrum of $S(k, \omega)$ \Ref{eq.22} allows investigations of
collective excitations in liquid cesium to be made in details.
Our analysis show, that the position of collective excitation
peak $\omega_{c}(k)$ in spectrum $S(k, \omega)$  depends on the
combination of frequencies $ \Omega_{1}^{2}$, $\Omega_{2}^{2}$,
$\Omega_{3}^{2}$ and $\Omega_{4}^{2}$. Our numerical
calculations demonstrate that $\omega_{c}(k)$ value is most sensitive
to relaxation frequency $\Omega_{2}^{2}$. On the other hand,
collective effects turn out to be connected with non-Markov processes in
liquids. 

In Fig.1 the comparison of our theory (solid line) and 
experimental data (circles) \cite{Boden.} for $S(k, \omega)$ for liquid
cesium at $T=380$ K is shown. From Fig. 1 it  is evident that 
our theory  absolutely agrees with  the experiment inthe whole range of
values of wave vector $k$. It is possible to see the good qualitative coincidence between
 the experiment and our theory from Fig. 2. Here 
dispersion of the frequency of collective excitations $\omega_{c}(k)$,
obtained from the  position of lateral peaks (the points present experimental
data; the  circles are our theoretical values), is presented.
 We calculated the first
four points in a statistical spectrum of the frequency dependent non-Markovity
parameter  $ \epsilon_{i}(k, \omega)$ for the whole range  of values 
of the wave vector $k$ . The results
of such calculations are shown in Fig. 3. In our opinion,the parameter
$\epsilon_{1}(k, \omega)$ represents special interest. From Fig. 3 it is 
apparent, that this parameter has maxima on frequencies that  coincide
with collective excitations in $S(k,\omega)$. 
The spectrum of 
$S(k,\omega)$ falls down smoothly  at large values 
of $k$, and the frequency dependent
non-Markovity parameter also  falls down smoothly to zero. Moreover,
"burst" of values $\epsilon_{1}(k,\omega)$ is observed on  common 
frequencies with $S(k,\omega)$. As indicated by  Fig. 3 
the non-Markovity parameter $\epsilon_{i}( \omega)$ for this
range of frequencies $\omega$ must satisfy  the condition
$\epsilon_{i}(k,\omega)>1$. The non-Markovity parameter 
$\epsilon_{1}(k,\omega)$ has
the distinct  expressed maximum on frequencies, relevant to collective
excitations. Proceeding from the received results, it is
reasonable to speak about the non-Markov nature of collective excitations in
liquid  cesium. This important
conclusion must be  taken into account while constructing  the
appropriate theories. 
The behavior of $\epsilon_{i}(k,\omega)$
for levels $i=2,3,4$ is  very interesting here. Further calculations
showes that the relationship $\epsilon_{3}(k,\omega) \approx 1$ takes place
for the high k-value range $1.15$ $\ang^{-1} \leq k \leq 2.55$ $\ang^{-1}$ and
for all frequencies $\omega$. Because of this, the more simple
approximation $M_{3}(t) 
\approx M_{2}(t)$ can be applied for this case. Moreover, there are 
cases, in which $\epsilon_{2}(k,\omega) \approx 1$ for  all $\omega$. 
It is possible to assume, that   good agreement with
the experiment gives correlation approximation for the memory function of the
second order $M_{2}(t) \approx M_{1}(t)$ in these cases.

Thus, the frequency dependent 
non-Markovity parameter introduced in Ref.\cite{Yulmetyev1}
$\epsilon_{i}(k,\omega)$ allows to reveal two important
features. First, its behavior allows to judge about the  properties of
non-Markovity in the whole  range of frequency spectrum and at various 
values of the wave vector $k$. Then, with its help if is possible to
judge about applicability of correlation approximation of the junior
order, for which 
less  difficult calculations are required.

The results of this Brief Report can be summarized as follow: (i)
$S(k,\omega)$ for liquid  cesium is found 
on the basis of the hypothesis of  time-scale
invariance of relaxation processes in liquids.
We have assumed, that  relaxation times of memory functions
$M_{4}(t)$ and $M_{3}(t)$ are approximately equal in this case and
relaxation time scales are invariant. Therefore, we have used correlation
approximation 
for memory function of the fourth order $M_{4}(t)$  for closing the chain of
the connected kinetic equations \Ref{chain}. As a result  we have
received  good agreement with  the experiment for all values of  wave
vector $k$ ($0.4$ $\ang^{-1} \leq k \leq 2.55$ $\ang^{-1}$).
(ii) For the  estimation of the received results the frequency - dependent
parameter 
of non-Markovity $\epsilon_{i}(\omega)$ ($i=1,2,3$ and $4$) was calculated
for all the  values of the wave vector $k$. It turn out , that the first
point in 
spectrum of non-Markovity parameter $\epsilon_{1}(\omega)$ has frequency
dependence similar to behavior the dynamic structure factor for liquid
cesium. The maximum of $\epsilon_{1}(k,\omega)$ and $S(k,\omega)$ for
values of the wave vector $0.4$ $\ang^{-1} \leq k \leq 2.55$ $\ang^{-1}$ 
appear at the same frequencies.
(iii) On the basis of 
our calculations we have established  the non-Markov nature of collective 
excitations at the microscopic level in liquid
cesium. It should be pointed out that the  developed approach is true 
especially for  the solution of
nonperturbative problems. Nevertheless, we believe that a systematic and
general theoretic approach, 
such as we have expounded, should play a useful role in analyzing and
classifying experimental data simulations and more elaborate models.
It 
is especially true for the  construction of  theories
describing high-frequency and short-range relaxation processes.
The calculated frequency parameter
$\epsilon_{1}(\omega)$ for every concrete value of  wave vector $k$
allows to describe the  effects of amplification or attenuation  of non-Markovity
within the whole frequency interval. From the preceeding  it is clear that the
idea of time-scale
invariance has large prospect for  the description of high and 
short-wave stochastic and relaxation processes in liquids. 

The authors acknowledge Dr. T. Scopigno, Professor A. G. Novikov, and Dr. R. K.
Sharma for useful discussion of INS and IXS data
for  liquid metals and Dr. L. O. Svirina for technical assistance. This 
work was partially supported by
the Russian Humanitarian Science Fund (Grant No. 00-06-00005a) and the
NIOKR RT Foundation [Grant No. 14-78/2000(f)].

\section {Figure caption}
Fig.1. Theoretical (solid line) and experimental \cite{Boden.} ($\circ$
$\circ$ $\circ$ $\circ$) values of dynamical structure factor
$S(k,\omega)$ for liquid cesium at $T=380 K$ at various wave-vectors $k =
\mid \bf k \mid$. 

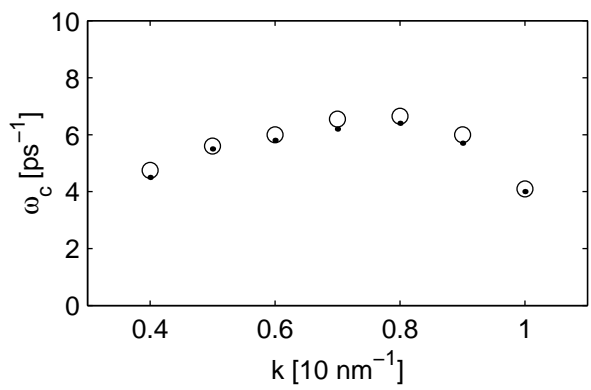
Fig.2. Comparison the values of frequency of collective
excitations in liquid cesium $\omega_{c}=\omega_{c}(k)$ at $T=380 K$ with
calculated values from experimental data \cite{Boden.} ($\cdot$ $\cdot$ $\cdot$
$\cdot$) and our theory ($\circ$ $\circ$ $\circ$ $\circ$).

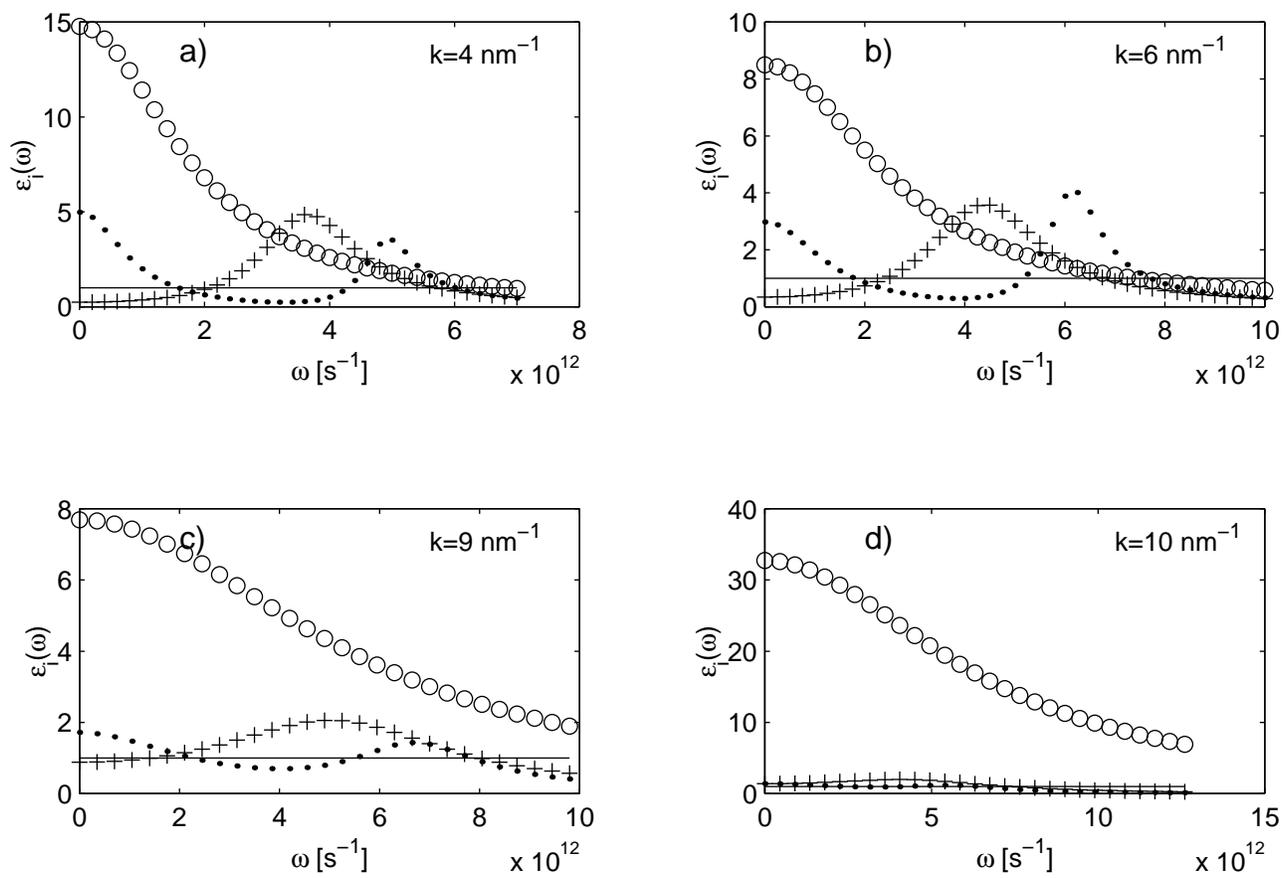
Fig.3. Frequency dependence of the first four points in statistical
spectrum of non-Markovity parameter $\epsilon_{i}=\epsilon_{i}(\omega)$;
($\cdot$ $\cdot$ $\cdot$ $\cdot$) correspond to $i=1$, ($+$ $+$ $+$ $+$)
relate to $i=2$, ($\circ$ $\circ$ $\circ$ $\circ$) present $i=3$, solid
line reflect values at $i=4$. In a wave vector range, $0.4 \ang^{-1} \leq
k \leq 1.1 \ang^{-1}$, collective excitations are exist.

\end {document}
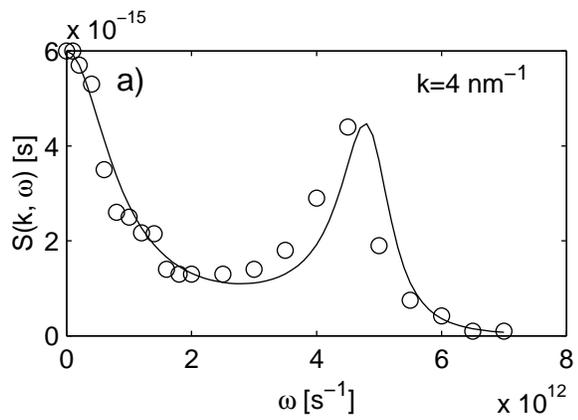
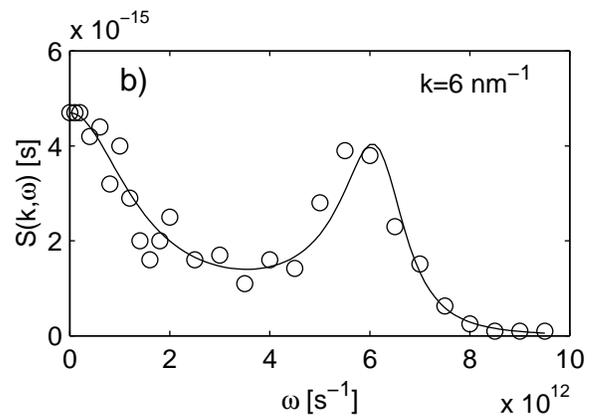
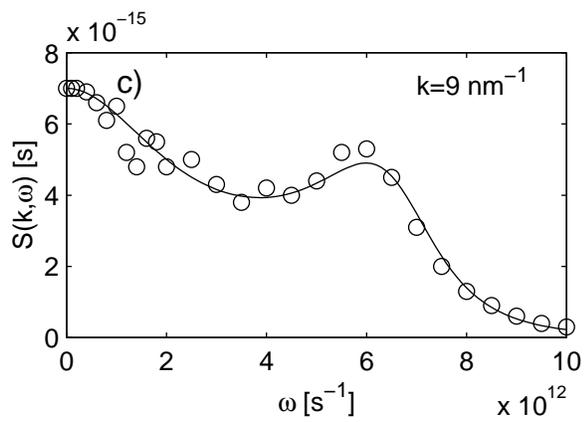
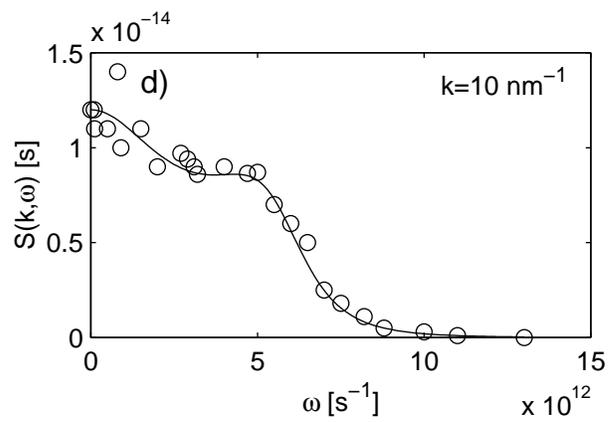
